\documentclass[aps,floatfix,twocolumn]{revtex4}

\usepackage{amsmath}
\usepackage{graphicx}
\usepackage{pslatex}
\usepackage{color}
\usepackage[normalem]{ulem}

\newcommand{\sys}{$\mathcal{S}$}
\newcommand{\env}{$\mathcal{E}$ }
\newcommand{\envnospace}{$\mathcal{E}$}
\newcommand{\sysspace}{$\mathcal{S}$ }

\begin{document}

\title{ thermalization of finite system and bath in quantum thermodynamics}

\title{ Simulating quantum thermodynamics of a finite system and bath with variable temperature}

\author{Phillip C. Lotshaw and Michael E. Kellman}

\affiliation{Department of Chemistry and Biochemistry and Institute of Theoretical Science, \\ University of Oregon \\ Eugene, OR 97403, USA}
\date{\today}

\begin{abstract}

We construct a finite bath with variable temperature for quantum thermodynamic simulations in which heat flows between a system \sysspace and the bath environment \env  in time evolution of an initial \sys\env pure state.  The bath consists of harmonic oscillators that are not necessarily identical.  Baths of various numbers of oscillators are considered; a  bath with five oscillators is used in the simulations.  The bath has a temperature-like level distribution.  This leads to definition of a system-environment microcanonical temperature $T_\mathcal{SE}(t)$ which varies with time.  The quantum state evolves toward an equilibrium state which is thermal-like, but there is significant deviation from the ordinary energy-temperature relation that holds for an infinite quantum bath, e.g. an infinite system of identical oscillators.  There are also deviations from the Einstein quantum heat capacity.  The temperature of the finite bath is systematically  greater for a given energy than the infinite bath temperature, and asymptotically approaches the latter as the number of oscillators increases.  It is suggested that realizations of these finite-size effects may be attained in computational and experimental dynamics of small molecules.

\end{abstract}

\maketitle

\section{introduction}

This paper considers computational simulation of a process of energy flow as a quantum system becomes entangled with a very small temperature bath.    In the corresponding ``classical" thermodynamic system, we would have an idea of a {\it variable temperature} as energy flows into the finite bath. Here we ask, does a simulacrum of thermodynamic behavior emerge when we make the bath very small?   Do reasonable ideas of a variable temperature hold, and  is there something akin to thermal equilibrium with a Boltzmann distribution?  We will find that with a very small ``thermal" environment, as small as five oscillators, it is possible to get behavior that is very much like thermodynamic behavior.  On the other hand, anomalies are observed related to the notion of temperature  with the small bath.   The work here builds on earlier simulations with a cruder, constant temperature bath \cite{polyadbath,deltasuniv,micro,Gemmerarticle,Olshanni,Belgians2}.  Questions of variable temperature in a very small quantum thermodynamic system and bath are of more than abstract interest.  Our simulations may not be too much simpler than what is called for in problems of practical import.  Quantum nanodevices can be imagined 
whose performance may depend on considerations similar to those here.     Similar in spirit to the approach taken here, quantum thermalization behavior of a pure quantum state has recently been observed experimentally in Bose-Einstein condensates containing as few as six-atoms \cite{Greiner2016}.  Recently \cite{Perez,Leitner2015,Leitner2018}, work on molecular ``quantum chaos" is being conceptualized as a venue for the exploration of contemporary ideas about the foundations of quantum thermodynamics, to which we turn next.

There have been a variety of simulations of quantum thermodynamic processes, including the very basic elementary process of heat flow into a bath \cite{polyadbath,deltasuniv,micro,Gemmerarticle,Olshanni,Belgians2}.  These have been successful in recovering standard thermodynamic behavior, with attainment of thermal equilibrium and a Boltzmann distribution for the system, with a properly behaving temperature.  However, these investigations have used rather simple models of the temperature bath, sometimes with a grossly discrete model of energy levels \cite{polyadbath,deltasuniv,Gemmerarticle}, in others with an approximation to continuous levels in the bath \cite{micro,Olshanni,Belgians2}, but always to our knowledge with a model of an effectively infinite bath with fixed temperature  in mind.    Usually also,  a very simple coupling between system and environment is  assumed, typically, a random matrix coupling without significant structure.  Paralleling (and sometimes preceding) these simulations, there has been a great deal of work 
\cite{deltasuniv,micro,Leitner2015,Leitner2018, Rigol, Deutsch, Deutsch1991, RigolReview,  tasaki1998, Gemmer:2009, Popescu2006, Popescu2009, Goldstein2006, vNcommentary, Goldstein2010, Goldstein2015, vNtrans, Reimann2008, Reimann2016,  Belgians, PolkovnikovicEntropy,  HanEntropy, KakEntropy, Reeb, Sun2014, Wolynes1990}  examining theoretical foundations of quantum thermodynamics.   Generally, this has focused on the large $N$ limit of quantum entangled systems. In our simulations here the focus is rather on the extent to which thermodynamic-like behavior persists as the total system becomes very small.   There have been  simulations examining ergodicity and energy flow in small total systems \cite{Leitner2015,Leitner2018,Rigol, Deutsch,Gruebele1995,Gruebele2003}, but these have not involved  the type of variable temperature analysis that is our focus here. We  construct a finite, variable temperature bath, also making use of  a structured coupling which is far more selective than the random matrix coupling used in many earlier simulations.  We will find that we can build a simulation model with  features very much like a variable temperature and thermalization, but with significant anomalies due to the finite bath, with some challenges to overcome having to do with the nature of the coupling.  

As noted briefly above, and in more detail in the concluding section, there are real molecular systems that could be considered as laboratories for ``post-classical" thermodynamic effects.   Consideration of small size is a recent ``dimension" of quantum thermodynamics beyond that introduced long ago with the advent of quantum levels.  A third innovation might come with novel effects from combining quantum time evolution with multiple small baths of the kind developed here for a single bath.

\section{Model System-Environment ``Universe"} \label{universe}

In this section, we detail the system and environment in our model;  we treat the system-environment interaction separately,  in Sections \ref{runawaysection}  and \ref{tamesection} .  

We will deal with a total system or ``universe" pure state for a coupled and entangled system and environment, or temperature bath.    The total Hamiltonian includes system \sys, environment \envnospace, and interaction \sys\env components

\begin{equation} \label{Htot} \hat H = \hat H_\mathcal{S} + \hat H_\mathcal{E} + \hat H_\mathcal{SE} \end{equation}

\noindent For the basis set we will use a truncation of the full \sys\env tensor product basis to a subset that contains all of the \sys\env basis states $\vert n \rangle \otimes \vert \epsilon\rangle$ in the energy range

\begin{equation} \label{basis} 0 \leq E_{n} + E_{\epsilon} \leq 13, \end{equation}

\noindent similar to the ``thermal basis" described in Ref. \cite{micro}.  The numerical convergence with this basis will be discussed in Section \ref{tamesection}.   Time evolution of the pure \sys\env state $\vert \Psi \rangle$ is carried out by numerically diagonalizing $\hat H$ and then calculating a series of timesteps using the Schr\"odinger equation $\vert \Psi(t) \rangle = \exp(-i\hat H t) \vert \Psi(0)\rangle$ ($\hbar=1)$. In this section we will develop the system and environment basis sets and Hamiltonians $\hat H_\mathcal{S}$ and $\hat H_\mathcal{E}$;  later sections develop   $\hat H_\mathcal{SE}$.   

The system Hamiltonian consists of a set of five evenly spaced  levels

\begin{equation} \langle n \vert \hat H_\mathcal{S} \vert n \rangle = \hbar \omega_\mathcal{S}n, \end{equation}

\noindent with frequency $\omega_\mathcal{S} =$  0.5 and quantum number $n = 0,1,...,4$. These choices of $\omega_\mathcal{S}$ and $n$ give a maximum system energy $ E_\mathcal{S}^{max} = 2$ that is reasonably small compared to the initial \sys\env state total energies we will consider in this paper $\langle \hat H \rangle \gtrsim 4$, where $\hat H$ is the total Hamiltonian of Eq. \ref{Htot}.  With larger  $ E_\mathcal{S}^{max}$ we have found that it is more difficult to get good system thermalization, since very few environment levels are paired with the highest energy system levels at the total energy $\langle \hat H \rangle$ when $ E_\mathcal{S}^{max} \approx \langle \hat H \rangle$.  This choice of $\omega_\mathcal{S}$ and $n= 0,1,...,4$ ensures that there is always a fair amount of energy in the environment, so that it can act properly as a heat bath to the system in our simulations.

We want to have an environment  or bath \env with certain properties more general than in earlier work \cite{polyadbath,deltasuniv,micro,Gemmerarticle,Olshanni,Belgians2}, and more similar to real physical systems.  We want the temperature to vary with energy, instead of being fixed.  We would also like for the energy and temperature to be close to proportional,  $T \sim E$, to the extent possible in a finite model, and exactly so in the limit of a large bath.  Furthermore, we may want the bath to have some significant structure, so that the couplings might also have some structure, unlike the abstract undefined environment levels with random couplings used earlier.  To do all of these things, we will construct the bath as a collection of oscillators.

Consider first a set of degenerate oscillators with equal frequencies and level spacings $\ \hbar\omega=1$.   This ``Einstein heat capacity" system has the well known degeneracy pattern and density of states

\begin{equation} \label{rhoEinstein} \rho_{Ein}(\eta,n_{tot}) = \frac{(\eta-1+n_{tot})!}{(\eta-1)!n_{tot}!},      \end{equation}

\noindent where $\rho_{Ein}(\eta,n_{tot})$ is the number of ways to distribute $n_{tot}$ total energy quanta into $\eta$ oscillators.  A more physically realistic model  will generalize to oscillators of different frequencies, so as to obtain something resembling a continuous distribution of levels, while  approximately maintaining the overall pattern of Eq. \ref{rhoEinstein}.  To this end, we will extend the distribution  $\rho_{Ein}$ to variable frequencies and energies using  a continuous function $\rho_{\mathcal{E}}$ that interpolates between the discrete points in Eq. \ref{rhoEinstein}.  Then, we will devise a set of distinct harmonic oscillator frequencies $\{\omega_{osc}\}$ that approximates the continuous distribution.  The total environment Hamiltonian is expressed as the sum of oscillator Hamiltonians

\begin{equation} \label{HE} \hat H_\mathcal{E} = \sum_{osc=1}^\eta \hat H_{osc}, \end{equation}

\noindent where the $\hat H_{osc}$ have energy eigenvalues

\begin{equation} \langle n_{osc} \vert \hat H_{osc} \vert n_{osc} \rangle =  \hbar \omega_{osc} n_{osc} , \end{equation}

\noindent where $n_{osc}$ is the quantum number of a given oscillator. We will analyze the density of states {$ \rho_{\hat H_\mathcal{E}}$} of the Hamiltonian $\hat H_\mathcal{E}$, finding good agreement with the continuous density $ \rho_{\mathcal{E}}$, and then analyze the temperature dependence of the model.  

\begin{figure}[h]
\begin{center}
\includegraphics[width=8cm,height=11cm,keepaspectratio,angle=0,trim=0 0 0 20]{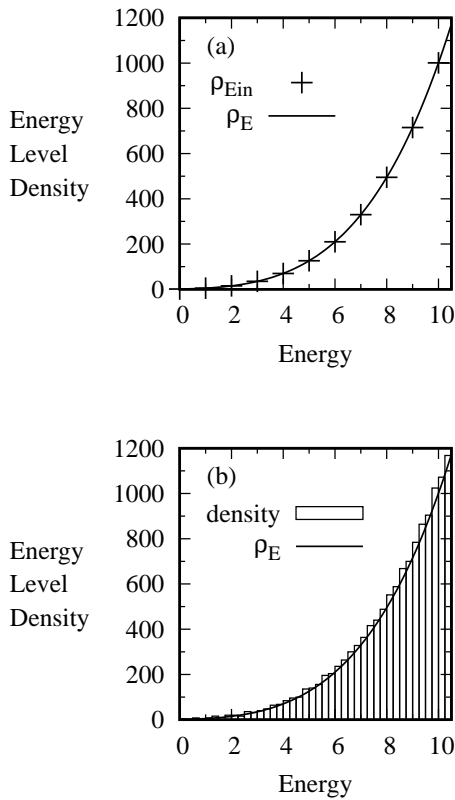}
\end{center}
\caption{(a) The continuous density $\rho_{\mathcal{E}}$ from Eq. \ref{rhoE} interpolates between the degenerate oscillator densities $\rho_{Ein}$ from Eq. \ref{rhoEinstein}.  (b) Oscillator density of states histogram for the five oscillator bath with the frequencies in Table \ref{frequency table}.}  
\label{density}
\end{figure}

We begin by developing a continuous density function $ \rho_{\mathcal{E}}$ in place of the highly degenerate density of Eq. \ref{rhoEinstein}.  The most straightforward way to do this is to replace the factorials in (\ref{rhoEinstein}) with Gamma functions 

\begin{equation} \label{rhoE} \rho_\mathcal{E}(E_\mathcal{E})= \frac{\Gamma(\eta+E_\mathcal{E})}{\Gamma(\eta)\Gamma(E_\mathcal{E}+1)},   \end{equation} 

\noindent where the discrete number of total quanta $n_{tot}$ has been replaced by a continuous environment energy $E_\mathcal{E}$.  The $\Gamma$ function extends the density to non-integer values of the energy $E_\mathcal{E}$, and agrees with the original density $\rho_{Ein}$ at integer $E_\mathcal{E}$ $= n_{tot}$, since for example $\Gamma(E_\mathcal{E} +1) = E_\mathcal{E}!$ $= n_{tot}!$ when $E_\mathcal{E} $  $= n_{tot}$ is an integer.  The top of Fig. \ref{density} shows how the continuous density $\rho_{\mathcal{E}}$ extends the degenerate oscillator density $\rho_{Ein}$ to non-integer $E_\mathcal{E} $.

 The next step is to devise a set of oscillator frequencies for the Hamiltonian $\hat H_\mathcal{E}$ in Eq. \ref{HE} with a density $\rho_{\hat H_\mathcal{E}}$ that follows the interpolating function  $\rho_{\mathcal{E}}$.  An $\eta=5$ oscillator bath will be used for the simulations.    This value of $\eta$ is large enough to give a density of states with an exponential-like dependence on energy, which will be imperative for Boltzmann thermalization of the system \sys, but also small enough to make the computations tractable.   The frequencies are generated as random numbers, to make the bath generic. 
We first tried generating random numbers $0.5 \leq \hbar\omega_{osc} \leq 1.5$ then rescaling the \ $\hbar \omega_{osc}$ so that their average was the same as the degenerate oscillator frequency $\hbar\omega=1$ seen in the top of Fig. \ref{density}.  However, when constructing the Hamiltonian $\hat H_\mathcal{E}$ in Eq. \ref{HE} using these frequencies, it was found that the resulting density of states $\rho_{\hat H_\mathcal{E}}$ was always  greater than the desired $\rho_\mathcal{E}$   of  Eq. \ref{rhoE}.  Instead, good agreement $\rho_{\hat H_\mathcal{E}}  \approx \rho_\mathcal{E}$ is consistently found by rescaling the random \ $\hbar\omega_{osc}$ values according to their geometric mean,

\begin{equation} \label{geometric mean}  \sqrt[\eta]{\prod_{osc=1}^\eta \hbar \omega_{osc}} = \hbar \omega = 1,  \end{equation} 

\noindent as discussed in detail shortly. {Eq. \ref{geometric mean} sets the unit of energy in this paper and also sets the relationship between the collection of variable frequencies $\{\hbar\omega_{osc}\}$ and the degenerate oscillator frequency $\hbar\omega$ assumed in connection with Eq. \ref{rhoEinstein}. The relation Eq. \ref{geometric mean} has previously been noted by Landau and Lifshitz \cite{LLStatPhyspp195} where it was also found to give the necessary link between variable and fixed frequency oscillators in a different context.

The $\hat H_\mathcal{E}$ that we use with Eq. \ref{HE} uses the frequencies given in Table \ref{frequency table} that come from randomly chosen values  that have been rescaled according to  Eq. \ref{geometric mean}.   The results are robust for other choices of random and rescaled $\{\hbar \omega_{osc}\}$. The density of states $\rho_{\hat H_{\mathcal{E}}}$ for this set of frequencies is shown in the histogram boxes in the bottom of Fig. \ref{density}, and is in excellent agreement with $\rho_\mathcal{E}$ of Eq.  \ref{rhoE}.  Recall that $\rho_\mathcal{E}$ also agrees with the fixed frequency $\rho_{Ein}$} as seen in the top of Fig. \ref{density}.  This demonstrates that Eq. \ref{geometric mean} gives the desired correspondence  between the densities of states for the variable and identical frequency oscillators: 

\begin{equation} \label{density equality}  \rho_{\hat H_\mathcal{E}} \approx \rho_\mathcal{E} = \rho_{Ein} \end{equation}

\noindent at integer energies $E_\mathcal{E} = n_{tot}$ and 

\begin{equation} \label{density equality 2} \rho_{\hat H_\mathcal{E}} \approx \rho_\mathcal{E}\end{equation}

\noindent  at non-integer energies (where the single-frequency $\rho_{Ein}$ is undefined in Eq. \ref{rhoEinstein}).  The correspondence between the somewhat random $\rho_{\hat H_\mathcal{E}}$   and the well-controlled, analytical $\rho_\mathcal{E}$ will allow us to determine analytical temperature relationships for our oscillator bath using the relatively simple function $\rho_\mathcal{E}$.  This is developed in the next section.

\begin{table}[h]
\begin{tabular}{|c|c|c|c|c|}
\hline
 $\hbar\omega_{1}$  &  $\hbar\omega_{2}$  &  $\hbar\omega_{3}$  &  $\hbar\omega_{4}$  &  $\hbar\omega_{5}$    \\
\hline
0.620\ 246& 0.735\ 401  & 1.146\ 315 & 1.316\ 886 & 1.453\ 415 \\
 \hline
\end{tabular}
\caption{Oscillator frequencies in the five harmonic oscillator environment shown to six decimal places.  }
\label{frequency table}
\end{table}

\section{Temperature}\label{temperature}

This rather involved section addresses key questions about the ``thermal" character introduced by the small finite bath in our model.  Does the standard infinite bath relation $E \sim T$ hold at high energy?  What is the low temperature behavior of the finite bath?  While sensible notions of temperature will emerge,  we will also see that there are anomalies in both of these aspects, related to the finite size of the bath.
    
We usually think of temperature in terms of a microcanonical ensemble with a very large, effectively infinite bath, so that the temperature is constant.  The temperature comes from the standard relation

\begin{equation} \frac{1}{T} = \frac{\partial S}{\partial E} \label{fundamentaltemp}   \end{equation}

\noindent applied to  the total system+environment  ${\mathcal{SE}}$ as the density of states is varied with energy.    In the situation envisaged in Fig. \ref{micro diagram}, we start by thinking instead of a temperature $T_\mathcal{E}$ for the bath environment initially in isolation from the system.   There are a multiplicity of initial separate system-bath combinations, each with the same total energy $E$;  an example is the red \sys\env state pair in the left of Fig. \ref{micro diagram}.  Each ${\mathcal{SE}}$ combination has its own initial system energy $E_\mathcal{S}$, bath energy $E_\mathcal{E}$,  and bath temperature $T_\mathcal{E}$.  The bath temperature $T_\mathcal{E}$ is based on a fixed $E_\mathcal{E}$ microcanonical energy that is defined only before the interaction with the system has begun -- the system in our simulations starts in a single zero-order state -- so there is no meaningful independent system temperature. Then, heat flows between system and  bath, leading to a finite change in a temperature that we want to be defined for the final equilibrium state, and perhaps in between as well.    The  final temperature $T_{\mathcal{SE}}$ after the heat flow comes from the microcanonical ensemble for the total system \sys\envnospace, which consists of the union of all the system-bath sub-ensembles, all with total \sys\env energy {$E$}, as in the right of Fig. \ref{micro diagram}.   An interesting relation Eq. \ref{TSE} will be found to hold between the inverse temperature $1/T_{\mathcal{SE}}$ of the complete ensemble of the \sys\env total system, and the average of the inverse temperatures $1/T_\mathcal{E}$ of the baths of the sub-ensembles.  In fact, it will be possible to define a time-varying ``master temperature" $T_\mathcal{SE}(t)$ in Eq. \ref{timedeptemp}  for the time-dependent intermediate state $\vert \Psi (t)\rangle$  in the equilibration process.  Thus,  we  will obtain a satisfying unified description of all the  possible processes of the type in Fig. \ref{micro diagram}.

\subsection{Temperature for Initial Isolated Environment} \label{TE section}

\begin{figure}
\begin{center}
\includegraphics[width=8.5cm,height=8.5cm,keepaspectratio,angle=0]{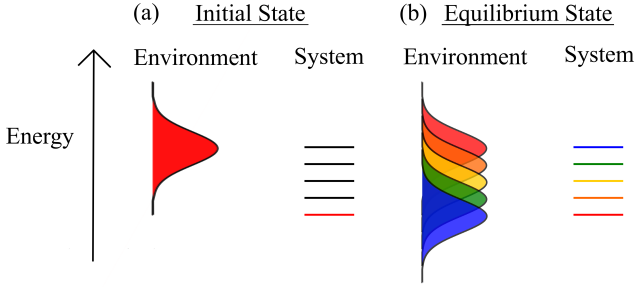}
\end{center}
\caption{(a)  Schematic example of an \sys\env initial state with the system in the lowest energy level and the environment in a high-energy Gaussian initial state as described in Section \ref{initial states}.  The temperature is $T_\mathcal{E}(E_\mathcal{E})$ from Eq. \ref{TE}.  (b)  Schematic of the same state after \sys\env equilibration, where now there is an \sys\env state pair for each system level, all at the same total \sysspace + \env energy.  The temperature is $ T_{\mathcal{SE}}$ from Eq. \ref{TSE}, which is the average of the $1/T_\mathcal{E}$ across all of the \sys\env state pairs. }
\label{micro diagram}
\end{figure}

First, we develop the temperature $T_\mathcal{E}$ for a finite environment that is thermally isolated from the system.   (This will turn out to be the initial state temperature in the time-dependent  temperature $T_\mathcal{SE}(t)$ to be developed in Section \ref{timedepT}.)   We will compare this finite bath to an infinite ``true" temperature bath of infinitely many oscillators.     The system is in a single zero-order initial state $n_0$, corresponding to our initial state in Fig. \ref{micro diagram}.  The total energy is $E$, the system has energy $E_\mathcal{S} = E_{n_0}$, and the environment has energy $E_\mathcal{E} = E - E_\mathcal{S}$. The temperature is defined using the standard thermodynamic relation of Eq. \ref{fundamentaltemp}.      This  is evaluated using the Boltzmann entropy $S = k_B \ln W(n_0,E)$, with $W(n_0,E)$  the number of \sys\env states $\vert n_0, \epsilon\rangle$ in a microcanonical energy shell $[E - \delta E/2, E + \delta E/2]$, again with the system in the level $n_0$.  
Since $n_0$ is  fixed,  $W(E) = \rho_{\mathcal{E}}(E_\mathcal{E})\delta E$ is just the number of environment states, where $\rho_\mathcal{E}$ in Eq. \ref{rhoE} is the smoothed continuous density function describing the density of discrete states in our Hamiltonian $\rho_{\hat H_\mathcal{E}}$, following Eqs. \ref{density equality} and \ref{density equality 2}.   The initial  temperature is then related only to the environment,  and we will label it $T_\mathcal{E}$, and rewrite it in terms of the density $\rho_{\mathcal{E}}$ as

\begin{equation} \label{TEdef} \frac{1}{T_\mathcal{E}} = \frac{d\rho_\mathcal{E}/dE_\mathcal{E}}{\rho_\mathcal{E}}. \end{equation}

\noindent Using Eq. \ref{rhoE} for $\rho_\mathcal{E}$ then gives

\begin{equation} \label{TE} \frac{1}{T_\mathcal{E}} = \psi(E_\mathcal{E}+\eta) - \psi(E_\mathcal{E} +1) = \sum_{m=1}^{\eta-1} \frac{1}{E_\mathcal{E} +m} , \end{equation}

\noindent where $\psi(x) = (d\Gamma(x)/dx)/\Gamma(x)$ is the digamma function. The last equality comes analytically from $\eta-1$ applications of the recurrence relation  \cite{wolframDiGamma} $\psi(x) = \psi(x-1) + 1/(x-1)$ to the term $\psi(E_\mathcal{E} + \eta)$.   

{It is not  clear just from looking at Eq. \ref{TE} how our temperature $T_\mathcal{E}$ for the finite bath will behave in comparison to standard temperature-energy relations involving an infinite fixed-temperature bath.   In the next two subsections we will make this comparison, using the paradigmatic standard of an average oscillator in an infinite oscillator bath.  Section \ref{terelation} will discuss the convergence of $T_\mathcal{E}$ from Eq. \ref{TE} to the standard temperature-energy relation as the size of the bath is increased, with convergence to the high energy relation $T \sim E$.  Section \ref{anomalous} will discuss deviations related to the finite size of the bath, including deviations from $T=0$ at low energy, and deviations in the heat capacity even at high energy.}

\subsubsection{Comparison of finite and infinite bath: {energy-temperature relation} }  \label{terelation}

The heat bath described above is a finite collection of oscillators.    We will compare this to a true temperature bath consisting of an infinite collection of oscillators.  For this, we use the energy-temperature relation from Einstein and Planck  for a  harmonic oscillator in an infinite temperature bath:   

 \begin{equation} \label{EEinstein}  \langle n_{ osc} \rangle  = \frac{1}{e^{1/T}-1}  \end{equation}
 
 \noindent    ($\hbar\omega =1$ and  $k_B = 1$) , where $\langle n_{osc}\rangle$ is the expected number of energy quanta in the oscillator.  (This relation   was  obtained by Einstein in his heat capacity model \cite{EinsteinCollectedPapers1907} by treating a solid as a collection of identical oscillators in an exterior temperature bath using the canonical ensemble.   The result  is the same regardless of the ensemble setup, microcanonical or canonical.)     We will find that our $T_\mathcal{E}$ for the finite bath behaves much like a standard temperature, but also has significant differences from the Einstein relation Eq. \ref{EEinstein}, leading also to deviations in the heat capacity from the Einstein model.  However, we also find that $T_\mathcal{E}$ agrees properly with Eq. \ref{EEinstein} in the limit of a large number of oscillators.    
The development is based on  
the correspondence $\rho_\mathcal{E} \approx \rho_{\hat H_\mathcal{E}}$ in Eqs. \ref{density equality} and \ref{density equality 2}, recalling the remarks there about the analytical function  $\rho_\mathcal{E}$

\begin{figure}[h]
\begin{center}
\includegraphics[width=9cm,height=9cm,keepaspectratio,angle=0]{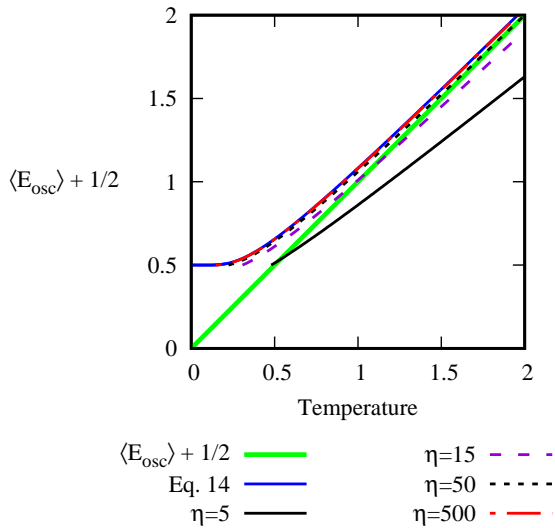}
\end{center}
\caption{Temperatures $T_\mathcal{E}$ converge to the Einstein solid temperature relation as the number of bath oscillators $\eta \to \infty$.  Deviations outside this limit are due to the finite size of the bath.  }
\label{templimit}
\end{figure}

These relationships are represented in Fig. \ref{templimit} and later for the heat capacity in Fig. \ref{heatcapacity}.  It will be instructive to consider the total energy of the ``Einstein oscillator" including both energy quanta and the zero-point energy, $\langle E_{osc}^{(+zp)} \rangle = \langle n_{osc} \rangle + 1/2$.  The blue curve in Fig. \ref{templimit} shows the relationship between $\langle E_{osc}^{(+zp)} \rangle$ and temperature based on Eq. \ref{EEinstein}. The curve begins at the zero-point energy at $T=0$, then quickly approaches the well-known quantum equipartition relation   

   \begin{equation} \label{TEinsteinlimit} \lim_{\langle n_{ osc} \rangle \to \infty} T = \langle n_{ osc} \rangle + \frac{1}{2} = \langle E_{osc}^{(+zp)} \rangle, \end{equation}
  \noindent shown by the green line in the background of the figure.

For comparison, Fig. \ref{templimit} also shows the relationship between $\langle E_{osc} \rangle + 1/2$ and $T_\mathcal{E}$ for finite oscillator baths with various $\eta$, again, based on the correspondence $\rho_\mathcal{E} \approx \rho_{\hat H_\mathcal{E}}$ in Eqs. \ref{density equality} and \ref{density equality 2}.    The average energy per oscillator from energy quanta $\langle E_{osc}\rangle \equiv E_\mathcal{E}/\eta$ is the analog for our bath of $\langle n_{osc}\rangle$ for the Einstein oscillator in Eqs. \ref{EEinstein} and \ref{TEinsteinlimit}.  The quantity 1/2 then shifts this up by the Einstein oscillator zero-point energy to allow for a direct comparison in the figure between our $T_\mathcal{E}$ and the temperature in the Einstein model.  In general, the exact zero-point energy in our model will not be 1/2 in our units (unlike the Einstein model), but will instead depend on the frequencies of the oscillators. Here, the $1/2$ is an arbitrary added quantity for the finite baths, inserted for comparison to the Einstein bath.

For the $\eta=5$ bath we use for our simulations,  shown by the black solid curve, the temperature behavior is significantly different than the blue infinite bath curve.   As we increase the number of oscillators $\eta$ we find that the curves get closer to the standard blue curve for an infinite bath.  For example, the dashed-double-dotted red line for $\eta=500$ oscillators rests on top of the blue line for the infinite bath $T$. 
The convergence towards Eq. \ref{EEinstein} with increasing $\eta$ confirms that our temperature gives the standard relation for an infinite bath in the thermodynamic limit $\eta\to\infty$, as expected with a reasonable temperature definition.  With this in mind, we next discuss in more detail the much more interesting question of anomalies in temperature behavior associated with small number of oscillators $\eta$ in the finite bath.

\subsubsection{{Anomalous temperature behavior associated with a very small bath}}   \label{anomalous}

The very small size of the $\eta=5$ bath leads to anomalous temperature behavior at both high and low energies, as seen in Fig. \ref{templimit}. 
First,  consider the behavior of $T_\mathcal{E}$ at low energies.  Recall that we treat this as a continuous variable that will be related to the continuous variable $E_E$ in Eq. \ref{TE}.    
The temperatures for all of the finite $\eta$ oscillator baths in Fig. \ref{templimit} are nonzero at the minimum  value of energy 1/2 in the figure (when $E_\mathcal{E} = 0$ in Eq. \ref{TE}, the rationale for the 1/2 being that given in the last subsection).  {The non-zero minimum temperatures}  seem  to be an unavoidable consequence of {combining a finite bath with  the standard temperature definition Eq. \ref{TEdef}.  The temperature is only zero when $d\rho_{\mathcal{E}}/dE_\mathcal{E} = \infty$ in Eq. \ref{TEdef} -- an  evidently impossible condition for a finite bath with a limited number of states.  However, as seen in Fig. \ref{templimit}, the curves for increasing $\eta$ converge to the standard infinite bath relation in which  $T=0$ at the minimum energy 1/2.

At high energy,  $T_\mathcal{E}$ approaches the asymptotic relation 

\begin{equation} \label{TElimit} \lim_{E_\mathcal{E}\to\infty} T_\mathcal{E} = \frac{E_\mathcal{E} + \eta/2}{\eta-1} = \left(\langle E_{osc}\rangle +\frac{1}{2}\right)\frac{\eta}{\eta-1}, \end{equation}

\noindent {where again $\langle E_{osc}\rangle = E_\mathcal{E}/\eta$   refers to the average energy per non-identical oscillator of the finite bath, although it also applies to an infinite ``Einstein bath" of identical oscillators.  Eq. \ref{TElimit} comes from the analytical limit of the right-hand side of Eq. \ref{TE}, which we evaluated using Mathematica.}     {Eq. \ref{TElimit} } {differs from the high-energy Einstein relation Eq. \ref{TEinsteinlimit} by the factor of $\eta/(\eta-1)$.  This difference is negligible in the thermodynamic limit $\eta\to\infty$ but very significant for small $\eta$, as seen by the differing slopes for the solid black and blue lines in Fig. \ref{templimit} at high energy.}

The differing slopes correspond to a difference in heat capacities

\begin{equation} \label{heatcapacity} C = \frac{d\langle E_{osc} \rangle}{dT} \end{equation}

   \begin{figure}[h]
\begin{center}
\includegraphics[width=9cm,height=9cm,keepaspectratio,angle=0]{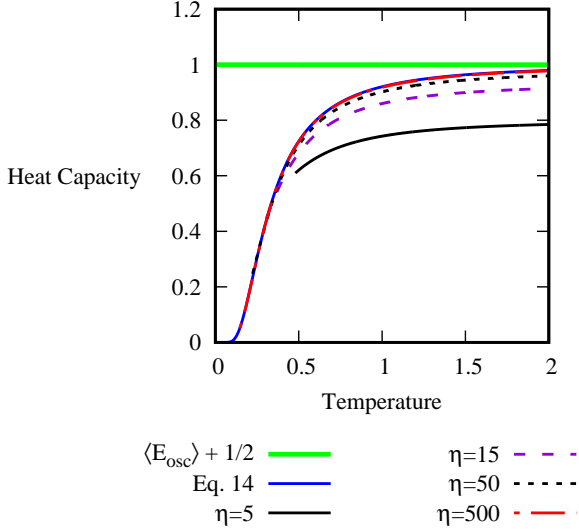}
\end{center}
\caption{ Heat capacities for the energy-temperature curves in Fig \ref{templimit}.}
\label{heatcapacity}
\end{figure}

\noindent {between the different temperature-energy relations.  The heat capacities for all of the temperature-energy curves in Fig. \ref{templimit} are plotted in Fig. \ref{heatcapacity}.  The heat capacity curves are similar to the standard Einstein behavior at low temperature,  but they are systematically lower at high temperature,   where they approach asymptotic values $C \to (\eta - 1)/\eta < 1$, less than both the Einstein relation and the standard equipartition result.

We will find in Section \ref{results on thermalization} that {the anomalous temperature behavior seen in Fig. \ref{templimit} is critical in obtaining the correct thermalized Boltzmann distribution for the system:  the anomalous scaling behavior $\sim \eta/(\eta-1)$ in the figure}  must be taken into account to correctly describe {the equilibrium \sysspace Boltzmann distribution and} the \sys\env thermodynamic behavior.

\subsection{System-Environment  Microcanonical Temperature}  \label{TSE section}

We  now consider the equilibrium \sys\env state and the temperature $T_{\mathcal{SE}}$ for the complex entangled state $\vert \Psi (t)\rangle$ shown schematically in the right of Fig. \ref{micro diagram}; this will be the equilibrium value of the time-dependent temperature $T_\mathcal{SE}(t)$ to be developed in Section \ref{timedepT}. 

 $T_\mathcal{SE}$ is defined following the same reasoning leading to Eq. \ref{TEdef}, giving

\begin{equation} \label{TSEdef} \frac{1}{T_{\mathcal{SE}}(E)} = \frac{d\rho_\mathcal{SE}/dE}{\rho_\mathcal{SE}}. \end{equation}

\noindent To evaluate the temperature we will examine  $\rho_\mathcal{SE}$ as the density of zero-order states, just as we did for the isolated bath temperature $\rho_\mathcal{E}$.  While there is some arbitrariness in doing this now with $\rho_\mathcal{SE}$, it is operationally simple, and seems at least as reasonable a choice as other possibilities.  It is consonant with what we have done with $\rho_\mathcal{E}$, and will lead to the simple result Eq. \ref{TSE}.  

The total density of \sys\env zero-order states at energy $E$ has contributions from all of the \sys\env state pairs that are in the microcanonical energy shell $E - \delta E/2 \leq E_\mathcal{S} + E_\mathcal{E} \leq E + \delta E/2$, {that is, each of the  \sys\env state pairs shown schematically in Fig. \ref{micro diagram}. } 
The total density of \sys\env states is the sum of bath densities that pair with each system level $n$ at the total energy $E = E_\mathcal{E} + E_{n}$,

\begin{equation} \label{rhoSE} \rho_\mathcal{SE}(E) = \sum_{n}\rho_{\mathcal{E}}(E-E_{n}).  \end{equation}

\noindent The \sys\env temperature can then be written as

\begin{equation} \frac{1}{T_{\mathcal{SE}}(E)} = \sum_{n} \frac{d\rho_{\mathcal{E}}(E-E_{n})/dE}{\sum_{m} \rho_{\mathcal{E}}(E-E_{m})}. \end{equation}

\noindent The derivatives can be rewritten in terms of $\rho_{\mathcal{E}}$ and $T_\mathcal{E}$ using Eq. \ref{TEdef}, giving

\begin{equation} \label{TSEmid} \frac{1}{T_{\mathcal{SE}}(E)} = \sum_{n} \frac{\rho_{\mathcal{E}}(E-E_{n})}{\sum_{m} \rho_{\mathcal{E}}(E-E_{m})}  \frac{1}{T_\mathcal{E}(E-E_{n})}. \end{equation} 

\noindent The fraction involving the densities gives the number of microcanonical states with the system in the level $E_{n}$ relative to the total number of microcanonical states.  This is simply the microcanonical probability of the system level $E_{n}$,

\begin{equation} \label{pmicro} \frac{\rho_{\mathcal{E}}(E-E_{n})}{\sum_{m} \rho_{\mathcal{E}}(E-E_{m})} = p_{micro}(E_{n}). \end{equation}

\noindent Putting this into Eq. \ref{TSEmid} gives the simple result

\begin{equation} \label{TSE} \frac{1}{T_{\mathcal{SE}}(E)} = \sum_{n}  \frac{p_{micro}(E_{n})}{T_\mathcal{E}(E-E_{n})} = \left\langle \frac{1}{T_\mathcal{E}(E-E_{n})} \right\rangle_{micro} \end{equation}

\noindent  Eq. \ref{TSE} says that the reciprocal temperature $1/T_\mathcal{SE}$ for the full \sys\env microcanonical ensemble is simply the average of the reciprocal environment temperatures $1/T_\mathcal{E}$ for each of the \sys\env state-pairs within the microcanonical ensemble.

It is interesting that the derivation of $T_\mathcal{SE}$ in Eqs. \ref{TSEdef}-\ref{TSE} used only the standard temperature definition in Eqs. \ref{TEdef} and \ref{TSEdef} and the choice of the zero-order basis for the densities of states $\rho_\mathcal{E}$ and $\rho_{\mathcal{SE}}$, used to formulate the sum in Eq. \ref{rhoSE}.  In this respect the relation Eq. \ref{TSE} is completely general, so it could also be used for other \sys\env thermodynamic models which could potentially be much different from the simple oscillator model we use here.

\subsection{Continuously varying time-dependent temperature} \label{timedepT}

The temperature relations in the previous sections were derived using the standard expression Eq. \ref{fundamentaltemp}  for the microcanonical ensemble, applied to the initial and final equilibrium states of the \sys\env universe.  It is useful  to consider a time-dependent generalization of the microcanonical temperature that can be defined {\it during} thermalization.    This uses  time-dependent system probabilities from the system reduced density operator $\hat \rho_\mathcal{S}(t)$ in place of the microcanonical probabilities in Eq. \ref{TSE}, giving 

\begin{equation} \label{timedeptemp} \frac{1}{T_{\mathcal{SE}}(E,t)} = \sum_{n}  \frac{\rho_\mathcal{S}^{{n},{n}}(t)}{T_\mathcal{E}(E-E_{n})} = \left\langle \frac{1}{T_\mathcal{E}(E-E_{n})} \right\rangle_{\hat \rho_\mathcal{S}(t)} \end{equation}

\noindent where $\rho_\mathcal{S}^{{n},{n}}$ is the probability of the system energy level $E_{n}$.    Note that this time-dependent temperature agrees with the initial temperature $T_\mathcal{E}$ in Eq. \ref{TE} and with the final temperature $T_{\mathcal{SE}}$ in Eq. \ref{TSE}. $T_\mathcal{SE}(t)$ is the ``master temperature" that describes the entire equilibration and thermalization process.  Using Eq. \ref{timedeptemp} we will be able to follow the time-dependent changes in temperature as \sysspace and \env begin in the initial state, exchange energy during thermalization, and eventually reach thermal equilibrium. This $T_{\mathcal{SE}} (t)$  is what we will be looking at as  the ``temperature" throughout the simulation.

\section{Initial states for the simulations} \label{initial states}

The calculations start at $t = 0$ with separable \sys\env initial states 

\begin{equation} \vert \Psi_{n_0}\rangle = \vert n_0 \rangle \vert \epsilon_0\rangle, \end{equation}

\noindent where the initial system level is $\vert n_0 \rangle$ and the initial environment state $\vert \epsilon_0\rangle$ has Gaussian distributed basis state probabilities

\begin{equation} \label{envstate} \vert \epsilon_0\rangle \sim \sum_\epsilon \exp\left({-\frac{(E_\epsilon-E_{\epsilon_0})^2}{2\sigma_\mathcal{E}^2}}\right) \vert \epsilon\rangle, \end{equation}

\noindent with $\sigma_\mathcal{E} = 0.5$ (the results are similar for other $0.1\leq \sigma_\mathcal{E}\leq 1$ that we have tested).  In Eq. \ref{envstate} the environment state is centered at an energy

\begin{equation} \label{Ecenter} E_{\epsilon_0} = E_0 - E_{n_0} \end{equation}

\noindent which varies with $n_0$, so that we are able to generate states that have the same nominal \sys\env} central energy $E_0 = E_{\epsilon_0} + E_{n_0}$ but different system levels $n_0$.  This will be useful for examining temperature equilibration,  where the final state in principle will depend on the total energy but not on $n_0$. 
An example of the total probability per unit energy for an $n_0=4$ initial state $\vert \Psi_{n_0}\rangle$ at energy $E_0 = 5$ is shown in the top of Fig. \ref{runaway}.  Each histogram bar in the figure shows the sum of \sys\env basis states probabilities within the surrounding zero-order energy unit; the actual state is naturally much more complex in the zero-order basis.  Note the logarithmic scale in the figure; the state is pretty sharply peaked around its nominal central energy.  A slight asymmetry can be observed about the central energy $E_0=5$.  This is because there are more basis states per unit energy above $E_0$ than below due to the increasing environment density of states. The asymmetry makes the average energy of the state slightly larger than the nominal energy $E_0$ in a way that depends on the environment density, which in turn depends on the environment energy $E_{\epsilon_0}$ and the system level $n_0$.  This gives a slightly different initial state energy for each $n_0$, but the energies are close to the same.

 We next consider the time evolution of this state, first with a random matrix coupling which we will find leads to pathological behavior, then with a more refined coupling that will be found to give physically satisfactory results.   
\begin{figure}[h]
\begin{center}
\includegraphics[width=6cm,height=15cm,keepaspectratio,angle=0,trim=0 0 0 20]{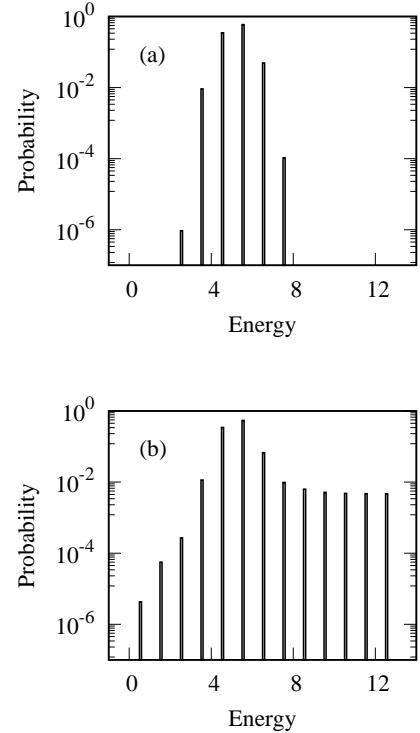}
\end{center}
\caption{Histogram of total quantum state probabilities per unit energy for an initial Gaussian state (a) and corresponding time-evolved equilibrium state (b) with a random matrix coupling with $k=0.0027$.   The total probability per unit energy does not converge to zero at high energy for the equilibrium state, indicating a problem with the coupling.  }
\label{runaway}
\end{figure}

\section{ random matrix coupling and runaway thermalization dynamics} \label{runawaysection}

In this section we begin developing the quantum dynamics with the coupling Hamiltonian $\hat H_\mathcal{SE}$ of  Eq. \ref{Htot}.  We begin with a standard type of coupling, the random matrix coupling,  used to model systems with classically chaotic dynamics \cite{Deutsch}, and often invoked   in  accounting for  the existence of thermalization in quantum thermodynamics \cite{Belgians2, Gemmerarticle, Deutsch}.  We used this in earlier simulations \cite{polyadbath,deltasuniv,micro} with good results.  However, we find here that with the introduction of a variable temperature, the random coupling introduces pathological behavior of runaway spreading of the wave packet.  Furthermore, the random coupling is a serious limitation in itself -- many important real systems do not have a random coupling.    Thus,  to understand thermalization for more realistic systems, we will want to explore more discriminating coupling forms.

The construction of $\hat H_\mathcal{SE}$ in Eq. \ref{Htot} as a random matrix coupling begins with a matrix $\hat R$ filled with off-diagonal elements 

\begin{equation}\label{random matrix} \langle n \vert \langle \epsilon \vert \hat R \vert \epsilon'\rangle \vert n'\rangle = R_{n\epsilon,n'\epsilon'}.\end{equation}

\noindent The $R_{n\epsilon,n'\epsilon'}$ are random complex numbers $R_{n\epsilon,n'\epsilon'} = X_{n\epsilon,n'\epsilon'} + iY_{n\epsilon,n'\epsilon'}$ as in Ref. \cite{Gemmerarticle}.  This is more generic than our previous work in Refs. \cite{polyadbath,deltasuniv,micro}, where we used real  $R_{n\epsilon,n'\epsilon'}$ to minimize numerical effort.  We generate the real and imaginary parts $X_{n\epsilon,n'\epsilon'}$ and  $Y_{n\epsilon,n'\epsilon'}$ each as random numbers   from a Gaussian distribution with standard deviation $\sigma=1$ with probabilities 

\begin{equation}  p(X_{n\epsilon,n'\epsilon'}) \sim e^{-X_{n\epsilon,n'\epsilon'}^2/2\sigma^2}, \end{equation}

\noindent  and similarly for the imaginary parts $Y_{n\epsilon,n'\epsilon'}$.  We set the diagonal elements to zero to preserve the oscillator energies in the zero-order basis, as was done previously in Ref. \cite{micro}.  The interaction Hamiltonian is then constructed by multiplying $\hat R$ by a parameter $k$ that sets the overall coupling strength, $\hat H_{SE} = k\hat R$.  This multiplication scales the random numbers so that their standard deviation becomes $\sigma=k$, consistent with the description in our earlier work \cite{polyadbath,deltasuniv,micro} (e.g. in Eq. 10 of Ref. \cite{polyadbath}).    We chose  $k$ to be the size of the average  level spacing of the system-environment universe at our initial state energy $E_0=5$, since we have found  that smaller $k$ do not give proper thermalization.

Fig. \ref{runaway} shows time evolution with this coupling.   With this coupling the initial Gaussian state associated with the  top panel evolves in time to the state of the bottom panel.   The time evolution evidently leads to runaway spreading of the wavepacket with  probability in high energy states that does not appear to be converging to zero.  This is not how a physically reasonable state should behave.

It is important to understand why this  coupling causes runaway behavior here, because it was not observed, at least so prominently, in our earlier simulations with a fixed temperature bath.  The coupling causes some spreading of the wavepacket to basis states of all energies, with the amount of probability per basis state decreasing rapidly as the states get farther off resonance from the initial state energy $E_0=5$.  This might seem to entail decreasing probabilities at the top edge of the basis.  However, the number of \env basis states per unit energy increases very rapidly with increasing energy in the variable temperature bath, as shown in Fig. \ref{density}, so that many more basis states contribute to the total probability in each successive energy unit.  Taken together, the total probability per unit energy doesn't converge to zero as it should, as clearly seen in Fig. \ref{runaway}.  This runaway coupling is a problem that needs to be addressed next.   

\section{Selective coupling ``tames" thermalization dynamics} \label{tamesection}

We will see that by defining a suitably much more selective coupling, physical results are obtained with both thermalization and contained spreading of the time-dependent quantum \sys\env state.   The basic idea is to ``tame" the coupling to limit the range of transitions, especially to high energy states.  

{As before with the random matrix coupling, we begin with a coupling constant $k$ and a random matrix $\hat R$ as in Eq. \ref{random matrix}. To construct $\hat H_\mathcal{SE}$, we take each individual matrix element of $k\hat R$ and multiply it by an exponential ``taming" factor that depends on the quantum number differences between the coupled states:}

\begin{equation}\label{tame coupling} \langle n \vert \langle \epsilon \vert \hat H_\mathcal{SE} \vert \epsilon'\rangle\vert n'\rangle = kR_{n\epsilon,n'\epsilon'}\exp\left({-\gamma_\mathcal{S}|\Delta n|-\gamma_\mathcal{E}\sum_{osc=1}^\eta|\Delta n_{osc}|}\right) \end{equation}

\noindent where $|\Delta n| = |n-n'|$ is the quantum number difference between the coupled system states and $\sum_{osc}|\Delta n_{osc}|$ is the total quantum number difference for the individual oscillators in the coupled environment states.  The parameters $\gamma_\mathcal{S}$ and $\gamma_\mathcal{E}$ suppress the coupling between \sys\env states depending on how much they vary in quantum number, for example the coupling that moves one quantum between the system and bath is stronger than the coupling that moves two quanta.  This limits the strength of transitions to high energy states, since they  typically differ significantly in their quantum number distributions,  thereby addressing the runaway problem. 

A coupling scheme similar to Eq. \ref{tame coupling} has been put forward by Gruebele \cite{Gruebele1995,Gruebele2003} in the context of intramolecular vibrational energy transfer, where he has argued that the exponential quantum-number dependence of the coupling is an approximate generic feature in molecular vibrational systems.  Deutsch \cite{Deutsch} has also said that a similar exponentially-tamed random matrix coupling can be obtained through a second-order perturbation theory analysis  and that the exponential taming is needed to prevent runaway behavior in large quantum thermodynamic systems. 

The tamed coupling has three parameters $k, \gamma_\mathcal{S},$ and $\gamma_\mathcal{E}$ that we choose somewhat arbitrarily for our model, with an aim towards obtaining physical thermalization behavior.  The $k$ sets the ``baseline" coupling strength; if $k$ is too small then thermalization will be impossible.  The $\gamma_\mathcal{E}$ restricts the \env transitions to address the runaway problem; it must be large enough to restrict the spreading with large energy differences, as needed for convergence, but also small enough to allow transfer between nearby \env levels, as needed for thermalization.  The $\gamma_\mathcal{S}$ controls how easily the system can transition between its levels; it must be small enough that all of the system levels can be accessed during the dynamics.

In our simulations we choose a coupling constant $k=0.15$.  This is much larger than the $k$ we used with the random matrix coupling, to balance the exponential taming factors. We choose a relatively small system taming factor $\gamma_\mathcal{S}=0.125$ and a large environment factor $\gamma_\mathcal{E} = 1$.  This parameter choice gives good system thermalization behavior while limiting the environment transitions strongly enough to get good convergence within our basis.  The effectiveness of this coupling and parameter choice is demonstrated by the time-evolved state in Fig. \ref{tame}.  The state corresponding to this figure began as an initial Gaussian state as seen in the top of Fig. \ref{runaway}, then it was evolved in time to equilibrium under the full Hamiltonian Eq. \ref{Htot} containing the tamed coupling interaction $\hat H_\mathcal{SE}$ from Eq. \ref{tame coupling}.   As seen in the histogram boxes in Fig. \ref{tame},  the total probability per unit energy is converging to zero at the top edge of the basis.   This shows that the tamed coupling has fixed the runaway problem of the random matrix coupling that was seen in the bottom of Fig. \ref{runaway}.  Using the tamed coupling we found good convergence with a maximum \sys\env energy $E^{max} = 13$ for the simulations in this paper.

\begin{figure}[h]
\begin{center}
\includegraphics[width=6cm,height=15cm,keepaspectratio,angle=0,trim=0 0 0 20]{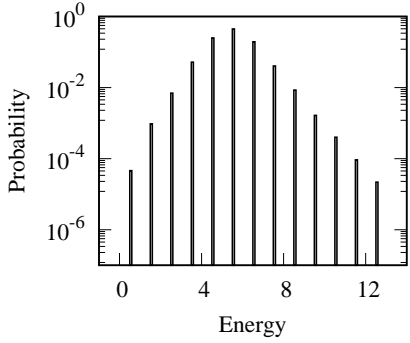}
\end{center}
\caption{Time evolved state with the ``tamed" coupling Eq. \ref{tame coupling} has probabilities that converge to zero at high energy.  The initial state was the same as panel (a) of Fig. \ref{runaway}.}
\label{tame}
\end{figure}

\section{Results:  Equilibration and Thermalization in the Simulations} \label{results on thermalization}

Now we examine  key aspects of the system dynamics during the approach to equilibrium:  behavior of the time-dependent temperature; and the question of equilibrated Boltzmann distribution with thermalization.   Is there thermodynamic-like behavior?  But do we also see  anomalous small-size temperature effects suggested by Fig. \ref{templimit}?

\subsection{Variable temperature and small-size effects}   \label{smallsize}

First we consider the computed time evolution of a set of initial states, constructed as described in Section \ref{initial states} with different initial system levels $n_0$ but the same nominal  energies $E_0=6$.  The  total energies for the various $n_0$ are somewhat larger,  as discussed in Section \ref{initial states},   with $6.116 \leq  \langle \hat H \rangle \leq 6.156$, where $\hat H$ is the total Hamiltonian Eq. \ref{Htot}. Taking $E =\langle \hat H \rangle$ in Eq. \ref{TSE} we  get for these states a narrow range of equilibrium {microcanonical} temperatures  $1.912 \leq T_\mathcal{SE}\leq 1.922$.  Roughly speaking, we can think of all the states as sharing the common energy $E \approx 6.14$, hopefully  corresponding in the simulations to a common final equilibrium temperature $T_\mathcal{SE} \approx 1.92$, where $1/T_\mathcal{SE}$ is the weighted average over all the initial state $1/T_\mathcal{E}$ at the common energy $E$, as in Eq. \ref{TSE}. We therefore test in the simulations whether the time-dependent temperature $T_\mathcal{SE}(t)$ of Eq. \ref{timedeptemp} equilibrates to the common temperature $T_\mathcal{SE} \approx 1.92$.

 Fig.  \ref{tempfig}  shows the time-dependent behavior of the temperatures $T_\mathcal{SE}(t)$ for each of the initial states $n_0$.  For each $n_0$, the temperature begins in its respective value for an isolated system and environment, $T_{\mathcal{SE}}(t=0) = T_\mathcal{E}$ (from Eqs. \ref{timedeptemp} and \ref{TE}).   Time evolution takes the temperatures to equilibrium, where they do  in fact fluctuate around the common approximate value $T_\mathcal{SE} \approx 1.92$. Thus, we are getting the common microcanonical $T_\mathcal{SE}$ value corresponding to energy $E \approx 6.14$, as  hoped for.  This result validates the path  of development in Section \ref{temperature}    regarding a variable temperature.    Observed small temperature fluctuations at equilibrium are due to the time-dependent fluctuations in the system density operator $\hat \rho_\mathcal{S}(t)$, whose behavior  will be discussed shortly in Section \ref{approach thermal eq}.

\begin{figure}
\begin{center}
\includegraphics[width=8.5cm,height=8.5cm,keepaspectratio,angle=0]{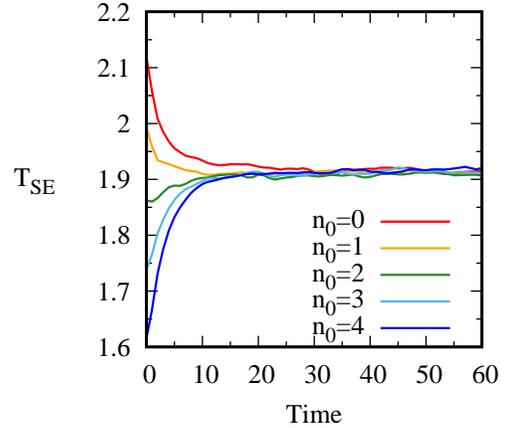}
\end{center}
\caption{Time-dependent temperatures $T_{\mathcal{SE}}(t)$ (Eq. \ref{timedeptemp}) for a series of calculations with approximately the same \sys\env  energy  $E \approx 6.14$ but different starting \sysspace levels $n_0$.  Each temperature evolves to approximately the same final temperature $T_{\mathcal{SE}} \approx 1.92$ from Eq. \ref{TSE}.}
\label{tempfig}
\end{figure}

It is a noteworthy prediction based on the considerations of Section  \ref{temperature} that the finite bath equilibrium temperatures in   Fig. \ref{tempfig} should be considerably higher than  would be expected using the infinite bath $T$ from Eq. \ref{EEinstein} based on the average number of quanta per degenerate oscillator {$\langle n_{osc}\rangle =$} $\langle E_{osc} \rangle$.  To test this, we calculated $\langle E_{osc}\rangle = \langle E_\mathcal{E}\rangle/\eta$ as the time-averaged equilibrium value for times $30 < t \leq 60$ averaged over  all of the simulations shown in Fig. \ref{tempfig}, giving $\langle E_{osc} \rangle = 1.117 \pm 0.004$.      The infinite bath limit temperature Eq. \ref{EEinstein} from this $\langle E_{osc}\rangle$ is $T = 1.564 \pm 0.004$, much smaller than our temperature $T_\mathcal{SE} = 1.92$.   This is because the finite bath temperatures $T_\mathcal{E}$ in Eq. \ref{TE} (which go into the calculation of the $T_\mathcal{SE}$ via Eq. \ref{TSE}) increase more rapidly with energy than the infinite bath $T,$ as was seen in Fig. \ref{templimit}.  Thus, the anomalous temperature scaling of the small environment is demonstrably evident from this analysis of Fig. \ref{tempfig}.  We will have more to say about the anomalous temperature in the next subsection.

\begin{figure}
\begin{center}
\includegraphics[width=8.5cm,height=8.5cm,keepaspectratio,angle=0]{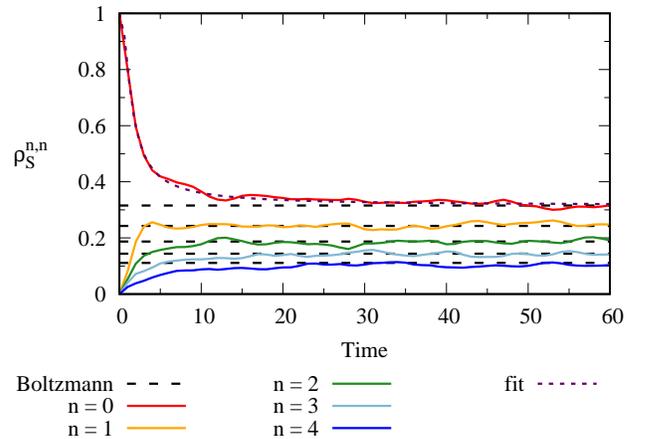}
\end{center}
\caption{System level probabilities  evolve in time to the  Boltzmann distribution at temperature $T_{\mathcal{SE}}(E=\langle \hat H \rangle)$ from Eq. \ref{TSE}.  {The decay of the initial state $n_0=0$ is described by Eq. \ref{power law} with $\tau = 1.02 \pm 0.03$ and $\delta = 2.38 \pm 0.06$.}}
\label{system probabilities}
\end{figure}

\subsection{Approach to thermal equilibrium and anomalous size effects} \label{approach thermal eq}

Next, we  consider {the behavior of the system in the approach to thermal equilibrium}.   Fig. \ref{system probabilities} shows an example of the time-dependent system probabilities $\rho_\mathcal{S}^{n,n}$ from the reduced density operator for an initial \sysspace level $n_0=0$  (the dynamics are similar for the other $n_0$).   As the state begins to evolve in time, much of the initial state probability is quickly lost to the other levels, followed by a much slower decay to the equilibrium Boltzmann distribution marked by the dotted lines.  The behavior can be fit by an empirical power law 

\begin{equation} \label{power law} \rho_\mathcal{S}^{n_0,n_0}(t) = \frac{1}{\sqrt{1+(t/\tau)^\delta}} \left( 1 - \frac{e^{-E_{n_0}/T_\mathcal{SE}}}{Z}\right) + \frac{e^{-E_{n_0}/T_\mathcal{SE}}}{Z} \end{equation}

\noindent  {where $\tau$ and $\delta$ are fit parameters and $\exp(-E_{n_0}/T_\mathcal{SE})/Z$ is the equilibrium Boltzmann probability at the temperature $T_\mathcal{SE}$, as will be discussed further shortly.   
Power law decays have been discussed by Gruebele \cite{Gruebele1998PNAS, Gruebele2003} as a generic feature in molecular vibrational systems that can be described by couplings similar to our Eq. \ref{tame coupling}.  The decay describes the nearly exponential drop of the initial state $n_0$ probability at short times and the longer decay to equilibrium. The other levels $n$ reach equilibrium at different timescales depending on how far they are from the initial level $n_0=0$, for example, $n=1$ reaches its equilibrium probability relatively quickly whereas it takes much longer for the $n=4$ level.  This stands in contrast to the dynamics under the simple random matrix coupling, where each system level evolves at approximately the same rate \cite{polyadbath}, without any sense of ``proximity" between nearby energy levels that facilitates their energy transfer.  Beyond simply being essential to converge the calculations, as discussed in Section \ref{tamesection},   it seems to us that the tamed coupling is also giving a much more realistic dynamics .

 \begin{figure*}
\begin{center}
\includegraphics[width=\textwidth,height=5cm,keepaspectratio]{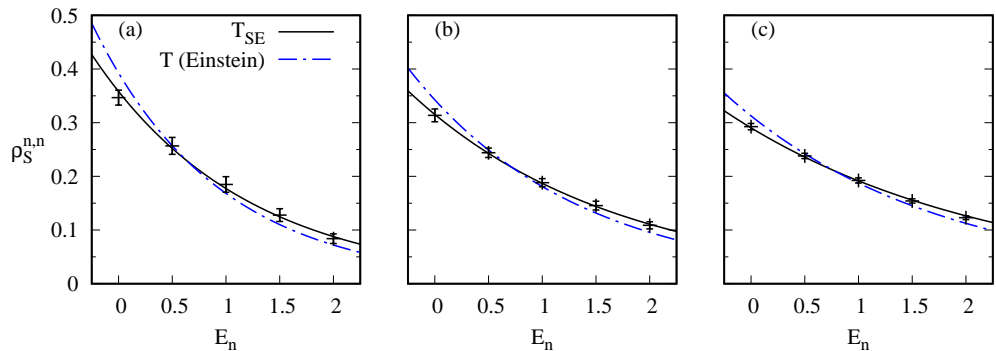}
\end{center}
\caption  {Time-averaged equilibrium system probabilities for three initial states (a), (b), and (c) with the energies and temperatures in Table \ref{temperature table}.  The Boltzmann distributions $\rho_\mathcal{S}^{n,n} \sim \exp({-E_n/T_\mathcal{SE}})$ at the analytical temperatures $T_\mathcal{SE}$ give very good descriptions of the system level probabilities $\rho_\mathcal{S}^{n,n}$, while the Boltzmann distributions at the infinite bath $T$  do not.  
}   

\label{boltzmann}
\end{figure*}

At long times, the system level probabilities fluctuate about a Boltzmann-appearing  distribution   $\rho_\mathcal{S}^{n,n}$ $\sim \exp(-E_n/T_{\mathcal{SE}})$ at the temperature $T_\mathcal{SE}$, shown as a black dotted line for each $E_n$.   The agreement with the Boltzmann distribution at $T_\mathcal{SE}$ is examined in Fig. \ref{boltzmann} across a range of initial state energies $E = \langle \hat H \rangle$ and corresponding temperatures listed in Table \ref{temperature table}.   The time-averaged system probabilities from the simulations are in very good agreement with the  analytical Boltzmann distributions at temperatures $T_\mathcal{SE}$ from Eq. \ref{TSE}.  For comparison, in Fig. \ref{boltzmann} we also show the Boltzmann distributions for the infinite bath temperatures $T$ calculated for the states, based on the average energy per bath oscillator observed in the simulations, see Table \ref{temperature table} and the discussion in the last paragraph of Section \ref{smallsize}.  The resulting temperatures are systematically lower than the $T_\mathcal{SE}$ values, and the corresponding Boltzmann distributions  do a poor job of describing the system probabilities.  Thus, the  observed thermalization to $T_\mathcal{SE}$  strongly reinforces that this is the correct thermodynamic temperature to describe the total system \sys\env. 

At this point it is appropriate to remark on the question of ``eigenstate thermalization" in our simulations.  The eigenstate thermalization hypothesis (ETH), that eigenstates of a suitable system-environment Hamiltonian reflect thermal properties \cite{Deutsch, Deutsch1991, RigolReview}, is widely regarded as an explanation for thermalization phenomena.    ETH is often justified through an appeal to chaotic dynamics of the kind that classically corresponds to a random matrix Hamiltonian. Chaotic dynamics become less certain the more that there is a ``taming" of the coupling, as used in this paper to get convergence of the dynamics, and     ETH thereby becomes less certain as well. Nonetheless, all of our initial states thermalize to their expected temperatures, and this is consistent with ETH.    In future work, we plan to explore the breakdown of ETH as reduced coupling strength makes questionable chaotic dynamics, ETH behavior, and thermalization itself.

Another point worth remark is alternatives to the random matrix-based couplings used in this paper.  Simple couplings based on linear combinations of raising and lowering operators are  used in many quantum thermodynamic investigations \cite{RigolReview}.  Accordingly, we have run calculations where we adopt a linear $k \hat x_i \hat x_j$ coupling. We find that this gives controlled spreading with semi-quantitative thermalization.  However, in comparison the  thermalization is significantly better with the random matrix tamed coupling calculations reported above.  The likely reason the random matrix works better for our setup is that our five-oscillator bath has approximate frequency resonances.  This is typical of many physical systems, e.g. a molecule embedded in a bath, which will almost inevitably have such ``anharmonic resonances."  A random coupling will better capture the effects of these resonances.  On the other hand, there are systems, e.g. of coupled bosons, where the $\hat x_i \hat x_j$ type coupling is more appropriate.  Based on our calculations, we believe that variable temperature baths can be devised appropriate to a variety of physical situations in ``tailor-made" fashion.

\begin{table}[h]
\begin{tabular}{|c|c|c|c|c|}
\hline
State & $E $  &  $T_\mathcal{SE}$  &  $\langle E_{osc} \rangle$  &  $T$ (Eq. \ref{EEinstein})    \\
\hline
 
(a) & 4.148 & 1.422 & 0.750 $\pm$ 0.005 & 1.180 $\pm$ 0.006\\
(b) & 6.118 & 1.913 & 1.121 $\pm$ 0.003 & 1.568 $\pm$ 0.003\\
 (c) & 8.099 & 2.406 & 1.499 $\pm$ 0.002 & 1.957 $\pm$ 0.002\\

 \hline
\end{tabular}
\caption{Energy and temperature data for Fig. \ref{boltzmann}.  The energies $E = \langle \hat H \rangle$ are from the full Hamiltonian in Eq. \ref{Htot} and the $T_{SE}(E)$ were calculated from Eq. \ref{TSE}.  The average bath-oscillator energies $\langle E_{osc}\rangle = E_\mathcal{E}/\eta$ were averaged over the same time window $30 < t \leq 60$ as the system probabilities in Fig. \ref{boltzmann} and the infinite bath $T$ were calculated from Eq. \ref{EEinstein} with $\langle n_{osc}\rangle = \langle E_{osc}\rangle$.}
\label{temperature table}
\end{table}

\section{summary and prospects}

This paper has considered a quantum description of energy flow from a system into a very small variable temperature bath.  We defined a system, consisting of a finite number of levels, and an environment,  consisting of  levels of a finite collection of harmonic oscillators (which constitutes the bath).  A set of identical oscillators was first considered, paralleling the Einstein heat capacity model.   To get something more like a continuous state distribution, we then took a collection of non-identical oscillators.  This gives a distribution of levels that closely tracks that of the bath of identical oscillators, but also has the desired feature of breaking the degeneracy, giving a quasi-continuous level distribution. The level pattern of this bath has a density of states that gives temperature-like behavior, using the standard statistical thermodynamic microcanonical relation  between temperature, energy, and density of states.  This defines the ``temperature" $T_\mathcal{E}$ for the finite bath.  This temperature differs significantly from that of the infinite oscillator bath, as seen in simulations with a bath with only $ \eta = 5$ oscillators.  We compared the energy-temperature relations for a single oscillator within the infinite bath (the well-known result of Einstein from his famous heat capacity paper) to the corresponding relation for a finite bath.    There are systematic differences, which are pronounced for $ \eta = 5$, and asymptotically approach the infinite bath at large $ \eta $.  The small bath has higher temperature for a given amount of energy per oscillator.  Very unlike the infinite bath, it also terminates at a temperature $T_\mathcal{E} > 0$, as seen in Fig. \ref{templimit}.  

Having devised the finite bath with temperature $T_\mathcal{E}$, we considered the process of heat flow from the system into this bath.   Simulations were performed of the process of heat flow to the finite bath in quantum time evolution.  First we used a random-matrix coupling of the kind that has been employed in many contexts, including successful quantum thermodynamic simulations \cite{Gemmerarticle,polyadbath,deltasuniv,micro}.  This however led to ``runaway spreading" of the quantum \sys\env wave function.  This is closely connected with the variable temperature of the bath -- a feature not present in earlier thermodynamic simulations.  The problem is that the density of states increases rapidly with increasing temperature, and the non-discriminate random coupling overpowers the quantum time evolution.  To solve this, we switched to a more selective coupling similar to the kind that has long been used \cite{Gruebele1995,Gruebele2003} in molecular simulations.  This selective coupling ``tames" the spreading of the wave function, so that runaway behavior is avoided.  The tamed coupling appears to be a realistic new feature needed to solve a real problem in the simulations.  

Next came computational examination of the temperature $T_\mathcal{SE}$ defined for the microcanonical ensemble of the \sys\env total system ``universe,"    including the time-dependent temperature $T_\mathcal{SE}(t)$ that varies continuously between the initial bath temperature $T_{\mathcal{E}}$ and the final \sys\env temperature $T_{\mathcal{SE}}$.    In simulations with the $ \eta = 5$ oscillator bath, starting with different initial system states but the same total system-environment energy, we tracked the temperature from its various initial values (because the bath has different energies depending on the system state) to its final value at equilibrium.  All the simulations went to essentially the same final temperature $T_\mathcal{SE}$, as desired.  The simulations with the bath of $ \eta = 5$ oscillators with selective coupling show equilibration to a Boltzmann-type distribution at the temperature $T_{\mathcal{SE}}$ implied by the initial energy of the total system.  As noted above, this temperature is markedly different from that of an infinite bath with the equivalent energy per bath oscillator.   In short, there are marked effects of the small finite bath on thermal behavior with variable temperature in the quantum simulations.

It is interesting to consider real situations in which to explore these finite size quantum thermodynamic effects.   Experiments on very small Bose-Einstein condensates, containing as few as six atoms \cite{Greiner2016}, may point the way to size-dependent variable temperature behavior similar to the oscillator model we have studied here.    Several investigators have proposed small molecules as laboratories for fundamental exploration of quantum thermodynamics and statistical mechanics.  Leitner \cite{Leitner2015,Leitner2018} has reviewed a method of using the eigenstate thermalization hypothesis to understand ergodicity and localization of energy within time-dependent molecular systems.    P\'erez and Arce \cite{Perez}  performed simulations of dynamics on a potential energy surface of the molecule OCS, which has a long history as an exemplar of problems of classically chaotic molecular dynamics.  They treat one of the vibrational modes of OCS as a ``system," and the other two modes as an ``environment," akin to what we do here, but with a two-mode bath that is much smaller even than what has been considered here.  They find a kind of thermalization of the system when it is excited with sufficient energy to have chaotic classical dynamics.  However, they did not engage in the kind of analytic treatment of temperature of the present paper.  If we go to a four-atom molecule, for example the important species C$_2$H$_2$ (acetylene) or H$_3$O$^+$ (hydronium ion), we could take as system one of the modes, e.g. a C-H stretch, leaving 5 vibrational modes as the bath, just as we do here.  This ignores rotational degrees of freedom; one could do experiments with angular momentum $J$ = 0; or alternately, allow $J$ excitations, which would become increasingly important at higher $J$, where rotation-vibration coupling would become important, giving the rotational degrees of freedom as a second bath or environment \envnospace$'$.    It is worth noting that molecular systems interacting with small baths are of interest in other contexts as well, e.g. in calculations of entanglement dynamics and spectroscopic signals \cite{Cina2014,Cina2017}.

As an alternative to the molecular dynamics simulations of Ref. \cite{Perez}, one could also use ``effective Hamiltonians" of the kind that have had vast use in molecular spectroscopy \cite{KellmanAnnRev,KellmanAccChemRes}.  It is notable that these Hamiltonians usually employ one or more ``polyad numbers" that constitute approximate constants of motion, valid on a limited time scale.  This makes these attractive systems in which to explore the effects of approximate constants as barriers to thermalization, a topic of considerable interest \cite{Deutsch} in contemporary theory of quantum thermodynamics.  The effective molecular polyad Hamiltonian can then be enhanced with polyad-breaking perturbations \cite{chakraborty,barnes2010,barnes2011} that correspond to real molecular dynamical effects.  These hierarchical dynamical systems could be ideal laboratories for investigation of thermodynamic processes on multiple time scales. 

As a final comment, taking a wider perspective on the work here, it may be worthwhile to consider that there are (at least) three dimensions of ``post-classical" effects in quantum thermodynamics.  The first of course is quantization of energy levels, introduced in the very beginnings of quantum physics by Planck in his black-body theory and  by Einstein in his famous heat capacity paper.  A second is finite size, as exemplified in this paper by the very small size (five oscillators) of the variable temperature bath.  A third involves quantum time evolution.  This might come with more complicated setups of finite size and time evolution than explored here.  One might consider  a system linking two baths of different sizes;  or a system linking two finite baths where the coupling of the system to each bath is different.  These would require far larger simulations than performed here.  We can readily imagine experimental realizations of these situations, e.g. with supramolecular arrangements of two or more molecules weakly linked by a third.      

\section{Acknowledgements}

P. L. thanks Rob Yelle and Craig Rasmussen for technical assistance with computations.  This work benefited from access to the University of Oregon high performance computer Talapas.

\bibliography{variabletame}

\begin{thebibliography}{45}
\expandafter\ifx\csname natexlab\endcsname\relax\def\natexlab#1{#1}\fi
\expandafter\ifx\csname bibnamefont\endcsname\relax
  \def\bibnamefont#1{#1}\fi
\expandafter\ifx\csname bibfnamefont\endcsname\relax
  \def\bibfnamefont#1{#1}\fi
\expandafter\ifx\csname citenamefont\endcsname\relax
  \def\citenamefont#1{#1}\fi
\expandafter\ifx\csname url\endcsname\relax
  \def\url#1{\texttt{#1}}\fi
\expandafter\ifx\csname urlprefix\endcsname\relax\def\urlprefix{URL }\fi
\providecommand{\bibinfo}[2]{#2}
\providecommand{\eprint}[2][]{\url{#2}}

\bibitem[{\citenamefont{Barnes and Kellman}(2013)}]{polyadbath}
\bibinfo{author}{\bibfnamefont{G.~L.} \bibnamefont{Barnes}} \bibnamefont{and}
  \bibinfo{author}{\bibfnamefont{M.~E.} \bibnamefont{Kellman}},
  \bibinfo{journal}{J. Chem. Phys.} \textbf{\bibinfo{volume}{139}},
  \bibinfo{pages}{21410893} (\bibinfo{year}{2013}).

\bibitem[{\citenamefont{Barnes et~al.}(2018)\citenamefont{Barnes, {Lotshaw},
  and {Kellman}}}]{deltasuniv}
\bibinfo{author}{\bibfnamefont{G.~L.} \bibnamefont{Barnes}},
  \bibinfo{author}{\bibfnamefont{P.~C.} \bibnamefont{{Lotshaw}}},
  \bibnamefont{and} \bibinfo{author}{\bibfnamefont{M.~E.}
  \bibnamefont{{Kellman}}}, \bibinfo{journal}{ArXiv e-prints}
  (\bibinfo{year}{2018}), \bibinfo{note}{arXiv:1511.06176},
  \eprint{1511.06176}.

\bibitem[{\citenamefont{Lotshaw and Kellman}(2019)}]{micro}
\bibinfo{author}{\bibfnamefont{P.~C.} \bibnamefont{Lotshaw}} \bibnamefont{and}
  \bibinfo{author}{\bibfnamefont{M.~E.} \bibnamefont{Kellman}},
  \bibinfo{journal}{J. Phys. Chem. A} \textbf{\bibinfo{volume}{123}},
  \bibinfo{pages}{831} (\bibinfo{year}{2019}).

\bibitem[{\citenamefont{Borowski et~al.}(2003)\citenamefont{Borowski, Gemmer,
  and Mahler}}]{Gemmerarticle}
\bibinfo{author}{\bibfnamefont{P.}~\bibnamefont{Borowski}},
  \bibinfo{author}{\bibfnamefont{J.}~\bibnamefont{Gemmer}}, \bibnamefont{and}
  \bibinfo{author}{\bibfnamefont{G.}~\bibnamefont{Mahler}},
  \bibinfo{journal}{Eur. Phys. J. B} \textbf{\bibinfo{volume}{35}},
  \bibinfo{pages}{255} (\bibinfo{year}{2003}).

\bibitem[{\citenamefont{Silvestri et~al.}(2014)\citenamefont{Silvestri, Jacobs,
  Dunjko, and Olshanii}}]{Olshanni}
\bibinfo{author}{\bibfnamefont{L.}~\bibnamefont{Silvestri}},
  \bibinfo{author}{\bibfnamefont{K.}~\bibnamefont{Jacobs}},
  \bibinfo{author}{\bibfnamefont{V.}~\bibnamefont{Dunjko}}, \bibnamefont{and}
  \bibinfo{author}{\bibfnamefont{M.}~\bibnamefont{Olshanii}},
  \bibinfo{journal}{Phys. Rev. E} \textbf{\bibinfo{volume}{89}},
  \bibinfo{pages}{042131} (\bibinfo{year}{2014}).

\bibitem[{\citenamefont{Esposito and Gaspard}(2003)}]{Belgians2}
\bibinfo{author}{\bibfnamefont{M.}~\bibnamefont{Esposito}} \bibnamefont{and}
  \bibinfo{author}{\bibfnamefont{P.}~\bibnamefont{Gaspard}},
  \bibinfo{journal}{Phys. Rev. E} \textbf{\bibinfo{volume}{68}},
  \bibinfo{pages}{066113} (\bibinfo{year}{2003}).

\bibitem[{\citenamefont{Kaufman et~al.}(2016)\citenamefont{Kaufman, Tai, Lukin,
  Rispoli, Schittko, Preiss, and Greiner}}]{Greiner2016}
\bibinfo{author}{\bibfnamefont{A.~M.} \bibnamefont{Kaufman}},
  \bibinfo{author}{\bibfnamefont{M.~E.} \bibnamefont{Tai}},
  \bibinfo{author}{\bibfnamefont{A.}~\bibnamefont{Lukin}},
  \bibinfo{author}{\bibfnamefont{M.}~\bibnamefont{Rispoli}},
  \bibinfo{author}{\bibfnamefont{R.}~\bibnamefont{Schittko}},
  \bibinfo{author}{\bibfnamefont{P.~M.} \bibnamefont{Preiss}},
  \bibnamefont{and} \bibinfo{author}{\bibfnamefont{M.}~\bibnamefont{Greiner}},
  \bibinfo{journal}{Science} \textbf{\bibinfo{volume}{353}},
  \bibinfo{pages}{794} (\bibinfo{year}{2016}).

\bibitem[{\citenamefont{P\'erez and Arce}(2018)}]{Perez}
\bibinfo{author}{\bibfnamefont{J.~B.} \bibnamefont{P\'erez}} \bibnamefont{and}
  \bibinfo{author}{\bibfnamefont{J.~C.} \bibnamefont{Arce}},
  \bibinfo{journal}{J. Chem. Phys.} \textbf{\bibinfo{volume}{148}},
  \bibinfo{pages}{214302} (\bibinfo{year}{2018}).

\bibitem[{\citenamefont{Leitner}(2015)}]{Leitner2015}
\bibinfo{author}{\bibfnamefont{D.~M.} \bibnamefont{Leitner}},
  \bibinfo{journal}{Adv. Phys.} \textbf{\bibinfo{volume}{64}},
  \bibinfo{pages}{445} (\bibinfo{year}{2015}).

\bibitem[{\citenamefont{Leitner}(2018)}]{Leitner2018}
\bibinfo{author}{\bibfnamefont{D.~M.} \bibnamefont{Leitner}},
  \bibinfo{journal}{Entropy} \textbf{\bibinfo{volume}{20}},
  \bibinfo{pages}{673} (\bibinfo{year}{2018}).

\bibitem[{\citenamefont{Rigol et~al.}(2008)\citenamefont{Rigol, Dunjko, and
  Olshanii}}]{Rigol}
\bibinfo{author}{\bibfnamefont{M.}~\bibnamefont{Rigol}},
  \bibinfo{author}{\bibfnamefont{V.}~\bibnamefont{Dunjko}}, \bibnamefont{and}
  \bibinfo{author}{\bibfnamefont{M.}~\bibnamefont{Olshanii}},
  \bibinfo{journal}{Nature} \textbf{\bibinfo{volume}{452}},
  \bibinfo{pages}{854} (\bibinfo{year}{2008}).

\bibitem[{\citenamefont{Deutsch}(2018)}]{Deutsch}
\bibinfo{author}{\bibfnamefont{J.~M.} \bibnamefont{Deutsch}},
  \bibinfo{journal}{Rep. Prog. Phys.} \textbf{\bibinfo{volume}{91}},
  \bibinfo{pages}{082001} (\bibinfo{year}{2018}).

\bibitem[{\citenamefont{Deutsch}(1991)}]{Deutsch1991}
\bibinfo{author}{\bibfnamefont{J.~M.} \bibnamefont{Deutsch}},
  \bibinfo{journal}{Phys. Rev. A} \textbf{\bibinfo{volume}{43}},
  \bibinfo{pages}{2046} (\bibinfo{year}{1991}).

\bibitem[{\citenamefont{D'Alessio et~al.}(2016)\citenamefont{D'Alessio, Kafri,
  Polkovnikov, and Rigol}}]{RigolReview}
\bibinfo{author}{\bibfnamefont{L.}~\bibnamefont{D'Alessio}},
  \bibinfo{author}{\bibfnamefont{Y.}~\bibnamefont{Kafri}},
  \bibinfo{author}{\bibfnamefont{A.}~\bibnamefont{Polkovnikov}},
  \bibnamefont{and} \bibinfo{author}{\bibfnamefont{M.}~\bibnamefont{Rigol}},
  \bibinfo{journal}{Advances in Physics} \textbf{\bibinfo{volume}{65}},
  \bibinfo{pages}{239} (\bibinfo{year}{2016}),
  \eprint{https://doi.org/10.1080/00018732.2016.1198134},
  \urlprefix\url{https://doi.org/10.1080/00018732.2016.1198134}.

\bibitem[{\citenamefont{Tasaki}(1998)}]{tasaki1998}
\bibinfo{author}{\bibfnamefont{H.}~\bibnamefont{Tasaki}},
  \bibinfo{journal}{Phys. Rev. Lett.} \textbf{\bibinfo{volume}{80}},
  \bibinfo{pages}{1373} (\bibinfo{year}{1998}).

\bibitem[{\citenamefont{Gemmer et~al.}(2009)\citenamefont{Gemmer, Michel, and
  Mahler}}]{Gemmer:2009}
\bibinfo{author}{\bibfnamefont{J.}~\bibnamefont{Gemmer}},
  \bibinfo{author}{\bibfnamefont{M.}~\bibnamefont{Michel}}, \bibnamefont{and}
  \bibinfo{author}{\bibfnamefont{G.}~\bibnamefont{Mahler}},
  \emph{\bibinfo{title}{Quantum Thermodynamics: Emergence of Thermodynamic
  Behavior Within Composite Quantum Systems (Second Edition)}}, Lecture Notes
  in Physics (\bibinfo{publisher}{Springer}, \bibinfo{year}{2009}).

\bibitem[{\citenamefont{Popescu et~al.}(2006)\citenamefont{Popescu, Short, and
  Winter}}]{Popescu2006}
\bibinfo{author}{\bibfnamefont{S.}~\bibnamefont{Popescu}},
  \bibinfo{author}{\bibfnamefont{A.~J.} \bibnamefont{Short}}, \bibnamefont{and}
  \bibinfo{author}{\bibfnamefont{A.}~\bibnamefont{Winter}},
  \bibinfo{journal}{Nature Phys.} \textbf{\bibinfo{volume}{2}},
  \bibinfo{pages}{754} (\bibinfo{year}{2006}).

\bibitem[{\citenamefont{Linden et~al.}(2009)\citenamefont{Linden, Popescu,
  Short, and Winter}}]{Popescu2009}
\bibinfo{author}{\bibfnamefont{N.}~\bibnamefont{Linden}},
  \bibinfo{author}{\bibfnamefont{S.}~\bibnamefont{Popescu}},
  \bibinfo{author}{\bibfnamefont{A.~J.} \bibnamefont{Short}}, \bibnamefont{and}
  \bibinfo{author}{\bibfnamefont{A.}~\bibnamefont{Winter}},
  \bibinfo{journal}{Phys. Rev. E} \textbf{\bibinfo{volume}{79}},
  \bibinfo{pages}{061103} (\bibinfo{year}{2009}).

\bibitem[{\citenamefont{Goldstein et~al.}(2006)\citenamefont{Goldstein,
  Lebowitz, Tumulka, and Zangh\`{i}}}]{Goldstein2006}
\bibinfo{author}{\bibfnamefont{S.}~\bibnamefont{Goldstein}},
  \bibinfo{author}{\bibfnamefont{J.~L.} \bibnamefont{Lebowitz}},
  \bibinfo{author}{\bibfnamefont{R.}~\bibnamefont{Tumulka}}, \bibnamefont{and}
  \bibinfo{author}{\bibfnamefont{N.}~\bibnamefont{Zangh\`{i}}},
  \bibinfo{journal}{Phys. Rev. Lett.} \textbf{\bibinfo{volume}{96}},
  \bibinfo{pages}{050403} (\bibinfo{year}{2006}).

\bibitem[{\citenamefont{Goldstein
  et~al.}(2010{\natexlab{a}})\citenamefont{Goldstein, Lebowitz, Tumulka, and
  Zangh\`{i}}}]{vNcommentary}
\bibinfo{author}{\bibfnamefont{S.}~\bibnamefont{Goldstein}},
  \bibinfo{author}{\bibfnamefont{J.~L.} \bibnamefont{Lebowitz}},
  \bibinfo{author}{\bibfnamefont{R.}~\bibnamefont{Tumulka}}, \bibnamefont{and}
  \bibinfo{author}{\bibfnamefont{N.}~\bibnamefont{Zangh\`{i}}},
  \bibinfo{journal}{Eur. Phys. J. H} \textbf{\bibinfo{volume}{35}},
  \bibinfo{pages}{173} (\bibinfo{year}{2010}{\natexlab{a}}).

\bibitem[{\citenamefont{Goldstein
  et~al.}(2010{\natexlab{b}})\citenamefont{Goldstein, Lebowitz, Mastrodonato,
  Tumulka, and Zangh\`{i}}}]{Goldstein2010}
\bibinfo{author}{\bibfnamefont{S.}~\bibnamefont{Goldstein}},
  \bibinfo{author}{\bibfnamefont{J.~L.} \bibnamefont{Lebowitz}},
  \bibinfo{author}{\bibfnamefont{C.}~\bibnamefont{Mastrodonato}},
  \bibinfo{author}{\bibfnamefont{R.}~\bibnamefont{Tumulka}}, \bibnamefont{and}
  \bibinfo{author}{\bibfnamefont{N.}~\bibnamefont{Zangh\`{i}}},
  \bibinfo{journal}{Phys. Rev. E} \textbf{\bibinfo{volume}{81}},
  \bibinfo{pages}{011109} (\bibinfo{year}{2010}{\natexlab{b}}).

\bibitem[{\citenamefont{Goldstein et~al.}(2015)\citenamefont{Goldstein, Hara,
  and Tasaki}}]{Goldstein2015}
\bibinfo{author}{\bibfnamefont{S.}~\bibnamefont{Goldstein}},
  \bibinfo{author}{\bibfnamefont{T.}~\bibnamefont{Hara}}, \bibnamefont{and}
  \bibinfo{author}{\bibfnamefont{H.}~\bibnamefont{Tasaki}},
  \bibinfo{journal}{New J. Phys.} \textbf{\bibinfo{volume}{17}},
  \bibinfo{pages}{045002} (\bibinfo{year}{2015}).

\bibitem[{\citenamefont{von Neumann}(2010)}]{vNtrans}
\bibinfo{author}{\bibfnamefont{J.}~\bibnamefont{von Neumann}},
  \bibinfo{journal}{Eur. Phys. J. H} \textbf{\bibinfo{volume}{35}},
  \bibinfo{pages}{201} (\bibinfo{year}{2010}), \bibinfo{note}{translated by
  Roderich Tumulka}.

\bibitem[{\citenamefont{Reimann}(2008)}]{Reimann2008}
\bibinfo{author}{\bibfnamefont{P.}~\bibnamefont{Reimann}},
  \bibinfo{journal}{Phys. Rev. Lett.} \textbf{\bibinfo{volume}{101}},
  \bibinfo{pages}{190403} (\bibinfo{year}{2008}).

\bibitem[{\citenamefont{Reimann}(2016)}]{Reimann2016}
\bibinfo{author}{\bibfnamefont{P.}~\bibnamefont{Reimann}},
  \bibinfo{journal}{Nature Comm.} \textbf{\bibinfo{volume}{7}},
  \bibinfo{pages}{10821} (\bibinfo{year}{2016}).

\bibitem[{\citenamefont{Esposito et~al.}(2010)\citenamefont{Esposito,
  Lindenberg, and den Broeck}}]{Belgians}
\bibinfo{author}{\bibfnamefont{M.}~\bibnamefont{Esposito}},
  \bibinfo{author}{\bibfnamefont{K.}~\bibnamefont{Lindenberg}},
  \bibnamefont{and} \bibinfo{author}{\bibfnamefont{C.~V.} \bibnamefont{den
  Broeck}}, \bibinfo{journal}{New J. Phys.} \textbf{\bibinfo{volume}{12}},
  \bibinfo{pages}{013013} (\bibinfo{year}{2010}).

\bibitem[{\citenamefont{Polkovnikov}(2011)}]{PolkovnikovicEntropy}
\bibinfo{author}{\bibfnamefont{A.}~\bibnamefont{Polkovnikov}},
  \bibinfo{journal}{Ann. of Phys.} \textbf{\bibinfo{volume}{326}},
  \bibinfo{pages}{486} (\bibinfo{year}{2011}).

\bibitem[{\citenamefont{Han and Wu}(2015)}]{HanEntropy}
\bibinfo{author}{\bibfnamefont{X.}~\bibnamefont{Han}} \bibnamefont{and}
  \bibinfo{author}{\bibfnamefont{B.}~\bibnamefont{Wu}}, \bibinfo{journal}{Phys.
  Rev. E} \textbf{\bibinfo{volume}{91}}, \bibinfo{pages}{062106}
  (\bibinfo{year}{2015}).

\bibitem[{\citenamefont{Kak}(2007)}]{KakEntropy}
\bibinfo{author}{\bibfnamefont{S.}~\bibnamefont{Kak}}, \bibinfo{journal}{Int.
  J. Theo. Phys.} \textbf{\bibinfo{volume}{46}}, \bibinfo{pages}{860}
  (\bibinfo{year}{2007}).

\bibitem[{\citenamefont{Reeb and Wolf}(2014)}]{Reeb}
\bibinfo{author}{\bibfnamefont{D.}~\bibnamefont{Reeb}} \bibnamefont{and}
  \bibinfo{author}{\bibfnamefont{M.~M.} \bibnamefont{Wolf}},
  \bibinfo{journal}{New J. Phys.} \textbf{\bibinfo{volume}{16}},
  \bibinfo{pages}{103011} (\bibinfo{year}{2014}).

\bibitem[{\citenamefont{Xu et~al.}(2014)\citenamefont{Xu, Li, Liu, and
  Sun}}]{Sun2014}
\bibinfo{author}{\bibfnamefont{D.~Z.} \bibnamefont{Xu}},
  \bibinfo{author}{\bibfnamefont{S.~W.} \bibnamefont{Li}},
  \bibinfo{author}{\bibfnamefont{X.~F.} \bibnamefont{Liu}}, \bibnamefont{and}
  \bibinfo{author}{\bibfnamefont{C.~P.} \bibnamefont{Sun}},
  \bibinfo{journal}{Phys. Rev. E} \textbf{\bibinfo{volume}{90}},
  \bibinfo{pages}{062125} (\bibinfo{year}{2014}).

\bibitem[{\citenamefont{Logan and Wolynes}(1990)}]{Wolynes1990}
\bibinfo{author}{\bibfnamefont{D.~E.} \bibnamefont{Logan}} \bibnamefont{and}
  \bibinfo{author}{\bibfnamefont{P.~G.} \bibnamefont{Wolynes}},
  \bibinfo{journal}{J. Chem. Phys.} \textbf{\bibinfo{volume}{93}},
  \bibinfo{pages}{4994} (\bibinfo{year}{1990}).

\bibitem[{\citenamefont{Bigwood and Gruebele}(1995)}]{Gruebele1995}
\bibinfo{author}{\bibfnamefont{R.}~\bibnamefont{Bigwood}} \bibnamefont{and}
  \bibinfo{author}{\bibfnamefont{M.}~\bibnamefont{Gruebele}},
  \bibinfo{journal}{Chem. Phys. Lett.} \textbf{\bibinfo{volume}{235}},
  \bibinfo{pages}{604} (\bibinfo{year}{1995}).

\bibitem[{\citenamefont{Gruebele}(2003)}]{Gruebele2003}
\bibinfo{author}{\bibfnamefont{M.}~\bibnamefont{Gruebele}},
  \bibinfo{journal}{Theor. Chem. Acc.} \textbf{\bibinfo{volume}{109}},
  \bibinfo{pages}{53} (\bibinfo{year}{2003}).

\bibitem[{\citenamefont{Landau and Lifshitz}(1980)}]{LLStatPhyspp195}
\bibinfo{author}{\bibfnamefont{L.~D.} \bibnamefont{Landau}} \bibnamefont{and}
  \bibinfo{author}{\bibfnamefont{E.~M.} \bibnamefont{Lifshitz}},
  \emph{\bibinfo{title}{Statistical Physics Part 1}}, Course on Theoretical
  Physics (\bibinfo{publisher}{Pergamon Press}, \bibinfo{year}{1980}),
  \bibinfo{edition}{3rd} ed., \bibinfo{note}{pp. 195-196}.

\bibitem[{wol()}]{wolframDiGamma}
\emph{\bibinfo{title}{Digamma function}},
  \bibinfo{howpublished}{\url{http://mathworld.wolfram.com/DigammaFunction.html}},
  \bibinfo{note}{accessed 4-30-2019}.

\bibitem[{\citenamefont{Einstein}(1989)}]{EinsteinCollectedPapers1907}
\bibinfo{author}{\bibfnamefont{A.}~\bibnamefont{Einstein}},
  \emph{\bibinfo{title}{``Planck's Theory of Radiation and the Theory of
  Specific Heat"}}, vol.~\bibinfo{volume}{2} of \emph{\bibinfo{series}{The
  collected papers of Albert Einstein}} (\bibinfo{publisher}{Princeton
  University Press}, \bibinfo{year}{1989}).

\bibitem[{\citenamefont{Gruebele}(1998)}]{Gruebele1998PNAS}
\bibinfo{author}{\bibfnamefont{M.}~\bibnamefont{Gruebele}},
  \bibinfo{journal}{Proc. Natl. Acad. Sci.} \textbf{\bibinfo{volume}{95}},
  \bibinfo{pages}{5965} (\bibinfo{year}{1998}).

\bibitem[{\citenamefont{Cheng and Cina}(2014)}]{Cina2014}
\bibinfo{author}{\bibfnamefont{X.}~\bibnamefont{Cheng}} \bibnamefont{and}
  \bibinfo{author}{\bibfnamefont{J.~A.} \bibnamefont{Cina}},
  \bibinfo{journal}{J. Chem. Phys.} \textbf{\bibinfo{volume}{141}},
  \bibinfo{pages}{034113} (\bibinfo{year}{2014}).

\bibitem[{\citenamefont{Kovac and Cina}(2017)}]{Cina2017}
\bibinfo{author}{\bibfnamefont{P.~A.} \bibnamefont{Kovac}} \bibnamefont{and}
  \bibinfo{author}{\bibfnamefont{J.~A.} \bibnamefont{Cina}},
  \bibinfo{journal}{J. Chem. Phys.} \textbf{\bibinfo{volume}{147}},
  \bibinfo{pages}{224112} (\bibinfo{year}{2017}).

\bibitem[{\citenamefont{Kellman}(1995)}]{KellmanAnnRev}
\bibinfo{author}{\bibfnamefont{M.~E.} \bibnamefont{Kellman}},
  \bibinfo{journal}{Ann. Rev. Phys. Chem.} \textbf{\bibinfo{volume}{46}},
  \bibinfo{pages}{395} (\bibinfo{year}{1995}).

\bibitem[{\citenamefont{Tyng and Kellman}(2007)}]{KellmanAccChemRes}
\bibinfo{author}{\bibfnamefont{V.}~\bibnamefont{Tyng}} \bibnamefont{and}
  \bibinfo{author}{\bibfnamefont{M.~E.} \bibnamefont{Kellman}},
  \bibinfo{journal}{Acc. Chem. Res.} \textbf{\bibinfo{volume}{40}},
  \bibinfo{pages}{243} (\bibinfo{year}{2007}).

\bibitem[{\citenamefont{Chakraborty and Kellman}(2008)}]{chakraborty}
\bibinfo{author}{\bibfnamefont{A.}~\bibnamefont{Chakraborty}} \bibnamefont{and}
  \bibinfo{author}{\bibfnamefont{M.~E.} \bibnamefont{Kellman}},
  \bibinfo{journal}{J. Chem. Phys.} \textbf{\bibinfo{volume}{129}},
  \bibinfo{pages}{171104} (\bibinfo{year}{2008}).

\bibitem[{\citenamefont{Barnes and Kellman}(2010)}]{barnes2010}
\bibinfo{author}{\bibfnamefont{G.~L.} \bibnamefont{Barnes}} \bibnamefont{and}
  \bibinfo{author}{\bibfnamefont{M.~E.} \bibnamefont{Kellman}},
  \bibinfo{journal}{J. Chem. Phys.} \textbf{\bibinfo{volume}{133}},
  \bibinfo{pages}{101105} (\bibinfo{year}{2010}).

\bibitem[{\citenamefont{Barnes and Kellman}(2011)}]{barnes2011}
\bibinfo{author}{\bibfnamefont{G.~L.} \bibnamefont{Barnes}} \bibnamefont{and}
  \bibinfo{author}{\bibfnamefont{M.~E.} \bibnamefont{Kellman}},
  \bibinfo{journal}{J. Chem. Phys.} \textbf{\bibinfo{volume}{134}},
  \bibinfo{pages}{074108} (\bibinfo{year}{2011}).

\end{thebibliography}

\end{document}